\documentstyle[preprint,aps,epsf]{revtex}  
\begin{document}
\draft
\preprint{TIT/HEP-245/COSMO-40}
\title{Compact Three Dimensional Black Hole:\\
Topology Change and Closed Timelike Curve}
\author{Masaru Siino\footnote{e-mail: msiino@phys.titech.ac.jp, 
JSPS fellow}}
\address{ Department of Physics, Tokyo Institute of Technology, Meguroku, 
Tokyo 152, Japan}
\date{\today}

\maketitle
\begin{abstract}
We present a compactified version of 
the 3-dimensional black hole recently found by considering extra identifications and 
determine the analytical continuation of the 
solution beyond its coordinate singularity by extending the identifications to the 
extended region of the spacetime. In the extended region of 
the spacetime, we find a topology change and non-trivial closed 
timelike curves both in the ordinary 3-dimensional black hole and in the 
compactified one. Especially, in the case of the compactified 3-dimensional black 
hole, we show an example of topology change from one double torus to eight spheres with 
three punctures.
\end{abstract}


\section{Introduction}
\hspace*{\parindent}
The topology change of the universe is a fascinating subject in general relativity.
Fujiwara, Higuchi, Hosoya, Mishima and the present author\cite{FHHMS} have recently found various quantum 
topology change solutions. Their quantum topology change solutions are
tunneling manifolds with the Euclidean signature. Such quantum effects 
occur only on the Planck mass scale.

In this paper we present a new topology change 
solution with the Lorentzian signature. It 
means that the topology change will occur in classical gravity.
There are some works treating the possibility of topology change  in 
classical gravity\cite{GST}. Especially, Geroch showed that in the case of a 
compact universe there must exist a closed timelike curve in the 
spacetime representing a topology changing process. In the same case, 
Tipler also pointed out that spacetime singularity necessarily appears in 
the spacetime if matter field satisfies the weak energy condition.
 
To study 
the topology change we must treat the complicated topology of spacetime. If we go down 
to the (2+1)-dimensional gravity, the problem becomes much simpler while the 
topology change is still 
non-trivial.  Further, the (2+1)-dimensional gravity is useful for studying 
roles of global structure of spacetime in classical and quantum 
gravity since there are no local dynamical degrees of freedom in the 
(2+1)-dimensional spacetime. Therefore we adopt the (2+1)-dimensional gravity as a 
good toy-model to study the Lorentzian topology change. 

To formulate the (2+1)-dimensional gravity, we consider the Teichm\"{u}ller 
parameters or moduli parameters as dynamical variables. Hosoya and 
Nakao, 
and Moncrief\cite{HNM} investigated a quantum dynamics of the Teichm\"{u}ller 
parameters of a torus universe in the ADM-formalism with the York timeslice.
The Teichm\"{u}ller parameters represent the identifications of points in 
the universal covering space of the torus and determines the shape of the 
torus. 
In the case of point particles, Deser, Jackiw and 'tHooft\cite{DJT} 
treated the mass of a point particle as 
a deficit angle of a Minkowski spacetime, which is determined by the 
change of period for the angular coordinate. 
We can see that in the (2+1)-dimensional gravity the identification of 
points of spacetime is crucial in the study of the dynamics of global 
structure.

Recently Ba\~{n}ados, Teitelboim and Zanelli\cite{BTZ} have found a 
(2+1)-dimensional black hole solution with a 
negative cosmological constant. Though the circular symmetric 
(2+1)-dimensional vacuum solution 
has no event horizon, their black hole solution constructed also by making 
certain identifications 
in the anti-de Sitter space has an event horizon. Since the Hawking 
temperature of the 3-dimensional black hole $T_H\propto\sqrt{M}$ becomes 
zero in the mass zero limit of the 3-dimensional black hole, it is 
expected that the Hawking radiation dies away when the mass of black 
hole becomes zero while it blows up in the 4-dimensional black 
hole. This suggests that the black hole evaporation can be treated 
semi-classically in (2+1)-dimensions. A 3-spacetime with a
negative cosmological 
constant has been studied since the variety of topology is 
rich in that space. Further, the violation of energy condition due to the 
negative cosmological constant allows topology 
change\cite{FHHMS}\cite{GST}.

In this paper, we will investigate the 3-dimensional black hole solution and 
its compactification, where the shape of the space 
is parametrized by a set of global parameters which becomes extra degrees of freedom.    
These 3-dimensional black hole solutions will be analytically extended. We analyze 
topology and the 
global causal structure of the whole of these spacetimes by studying the 
identification of points in the 3-anti-de Sitter spacetime. The similarity with the Misner 
solution suggests existence of closed timelike curves.

In the section 2, the 3-dimensional black hole solution is 
compactified to give a double torus space. The section 3 shows their 
analytical extensions. The topology is discussed there.
 The closed curves are analyzed in section 4. They make the global causal 
 structure clear. The final section is devoted to summary and discussions.

\section{Compactified 3-dimensional Black Hole}  
\hspace*{\parindent}

As shown by Ba\~{n}ados, Teitelboim and Zanelli\cite{BTZ}, there exists a 
black hole solution in the 
(2+1)-dimensional pure gravity with a negative cosmological constant. The 
metric of non-rotating black hole is given by
%
%
\begin{equation}
\label{eqn:3bh}
ds^2=-(r^2/l^2-M)dt^2+{dr^2 \over (r^2/l^2-M)}+r^2 d\phi^2 ,
\end{equation}
where $M$ is the mass of the black hole and $-1/l^2$ is equal to the cosmological 
constant. This solution is a maximally symmetric spacetime with a constant 
curvature $-2/l^2$ (3-anti-de 
Sitter) with certain identifications of points in spacetime. 
In this sence, this solution is regarded as the topological one. More 
precisely, we consider the identifications of points by a discrete subgroup 
of the isometry group $SO(2,2)$ acting on the 3-pseudo-sphere 
$x^2+y^2-z^2-w^2=l^2$, which is a 3-anti-de Sitter spacetime embedded 
into the flat spacetime with a signature $(--++)$.
The coordinate parametrization for the 3-dimensional black hole (3D-BH) in 
the external region  
($r\geq l$) is given as 
%
%
\begin{eqnarray*}
 ds^2&=& -dx^2-dy^2+dz^2+dw^2,\\
 x&=&\sqrt{r^2/M-l^2}\sinh {\sqrt{M} \over l}t,\\
 y&=& {r \over \sqrt{M}}\cosh\sqrt{M}\phi,\\
 z&=&\sqrt{r^2/M-l^2}\cosh {\sqrt{M} \over l}t,\\
 w&=& {r \over \sqrt{M}}\sinh \sqrt{M}\phi.
\end{eqnarray*}
The internal solution ($r<l$) is given by an analytical continuation of 
the expression written above as
%
%
\begin{eqnarray*}
x&=&\sqrt{l^2-r^2/M}\cosh {\sqrt{M} \over l}t,\\
y&=&{r \over \sqrt{M}}\cosh\sqrt{M}\phi,\\
z&=&\sqrt{l^2-r^2/M}\sinh {\sqrt{M} \over l}t,\\
w&=&{r \over \sqrt{M}}\sinh\sqrt{M}\phi.
\end{eqnarray*}
This chart of the 3D-BH solution covers the part of the pseudo-sphere in 
which $y^2-w^2>0$. In this region the periodic boundary condition of the 
angular coordinate, $(t, 
r, \phi) \leftrightarrow 
(t, r, \phi+2\pi)$, defines an identification of points and restricts its 
fundamental 
region. This identification is written in terms of $(x, y, z, w)$ as
%
%
\begin{equation}
\label{eqn:mat}
\begin{array}{l}
\left(\begin{array}{c}
x  \\
z
\end{array}\right) \longleftrightarrow 
\left(\begin{array}{c}
x  \\
z
\end{array}\right)\\
\left(\begin{array}{c}
y  \\
w
\end{array}\right)
\longleftrightarrow
\left(\begin{array}{cc}
\cosh 2\pi\sqrt{M} & \sinh 2\pi\sqrt{M} \\
\sinh 2\pi\sqrt{M} & \cosh 2\pi\sqrt{M} \\
\end{array}\right) 
\left(\begin{array}{c}
y  \\
w
\end{array}\right).
\end{array}
\end{equation}
This transformation is an element of the $SO(2,1)$ group, the subgroup of the 
isometry group $SO(2,2)$  of the 3-anti-de Sitter spacetime. We  
quotient the 3-anti-de Sitter spacetime by the discrete subgroup of 
$SO(2,2)$ 
generated by (\ref{eqn:mat}) to obtain the 3D-BH solution.
Considering also the coordinate parametrization for the Robertson-Walker 
chart of the anti-de Sitter 
spacetime, we find that this element of the group preserves not only the 
timeslice of the 3D-BH but also the timeslice of the RW-chart. Since each 
spatial hypersurface of the RW-chart of the 3-anti-de Sitter spacetime 
is a 2-hyperbolic space, we 
can construct various handle-body universes by certain tessellations on 
them so that the tessellation of the handle-body becomes consistent with 
the 3D-BH solution.
For example, we consider a double torus universe with 
a discrete subgroup of $SO(2,1)={\rm Isom}(H_2)$ generated by
%
%
\begin{equation}
\label{eqn:dt}
\begin{array}{l}
T_1=T ,\\
T_2=R^{-1}T R ,\\
T_3=R^{-2}T R^2 ,\\
T_4=R T R^{-1},
\end{array}
\end{equation}
where $T$ is a Lorentz boost and $R$ is a rotation by $\pi/4$.
The actions of these transformations on $H_2$ are shown in Fig.1.
In terms of ($x, y, z, w$), these $T$ and $R$ are represented by the 
matrices,
%
%
\begin{eqnarray}
\label{eqn:mdt}
T &=&
\left(\begin{array}{cccc}
1 & 0 & 0 & 0 \\
0 & \cosh\alpha & 0 & \sinh\alpha \\
0 & 0 & 1 & 0 \\
0 & \sinh\alpha & 0 & \cosh\alpha
\end{array}\right),\\
R &=&
\left(\begin{array}{cccc}
1 & 0 & 0 & 0 \\
0 & 1 & 0 & 0 \\
0 & 0 & \cos\pi/4 & -\sin\pi/4 \\
0 & 0 & \sin\pi/4 & \cos\pi/4
\end{array}\right).
\end{eqnarray}
%
%
\begin{figure}[h]
	\centerline{\epsfxsize=9.0cm \epsfbox{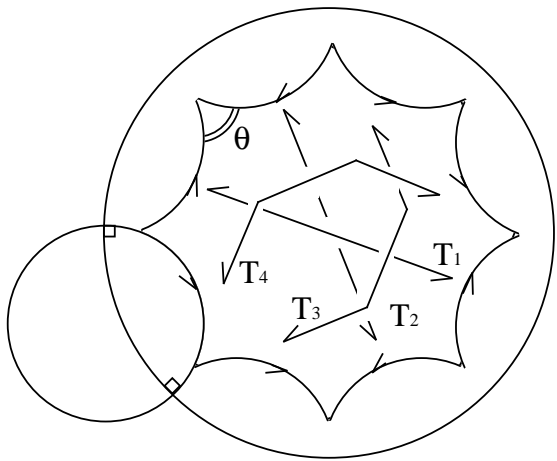}}
	\caption{An edge is identified with an edge in the opposite side.}
\end{figure}

Compared with (\ref{eqn:mat}) this double torus universe is consistent with 
the 3D-BH solution 
provided that $\alpha $ is equal to $2\pi\sqrt{M}$. To get a 
regular double torus (the angle $\theta$ in Fig.1 is $\pi/4$), 
however, $\alpha$ should be $2\tanh^{-1}\sqrt{(2\sqrt{2}-2)/3}$; otherwise 
the double torus has a conical singularity. From now on, we suppose that
%
%
\begin{equation}
2\pi\sqrt{M} = \alpha = 2 \tanh^{-1}\sqrt{{2 \sqrt{2} -2 \over 3}}\ \ ,
\end{equation}
for the regular double torus.
We call this regular double torus spacetime with a negative cosmological 
constant as {\it a compactified 3D-BH}. Because the double torus is constructed 
on 
the RW-chart, the foliation by the 2-hyperbolic spaces which are the 
universal covering spaces of the 
double tori covers only the region $y^2-z^2-w^2\geq 
0$. The 
compactified 3D-BH is the quotient of this region by 
the discrete subgroup. The compactified 3D-BH is 
singular only on a cone 
$y^2-z^2-w^2=0$ where the RW-chart is singular. The property of these singularities will be discussed in 
the 
next section.
\section{Singularity and Misner-like Extension}
\hspace*{\parindent}
In this section, we consider the causal structure of the 
3D-BH solution which is similar to that of the Misner spacetime. We
extend the 3D-BH spacetime in 
the same way as we normally treat the Misner solution\cite{MI}.
\subsection{Ordinary 3D-BH}
\hspace*{\parindent}
In the Misner solution the apparent singularity is merely a coordinate
singularity. We can extend the solution to the region beyond the 
coordinate singularity. It is well known that there are two possible extensions of 
the 
solution 
corresponding to the two directions along the spatial axis. To 
make all 
geodesics in these directions complete we should do both of the extensions.
The resultant spacetime, however, becomes non-Hausdorff at the boundary 
between the original region and the extended region as shown in\cite{HE}.

Now, we extend the 3D-BH by analogy with the Misner solution.
By a coordinate transformation $\rho=r^2$, the metric of the ordinary 
3D-BH (\ref{eqn:3bh}) changes to the following
form which has the Misner-like section:
%
%
\begin{equation}
\label{mis}
ds^2=\left\{-\left(M-\rho/l^2\right)^{-1}{d\rho^2 \over 4\rho}+\rho 
d\phi^2\right\}+\left(M-\rho/l^2\right) dt^2\ \ .
\end{equation}
When one approaches the singularity ($\rho=0$), this section behaves like 
the Misner solution. There are two extensions of spacetime, 
corresponding to the two directions along the $\phi$-axis, which are 
similar  to 
the extensions of the Misner solution. 
We consider the following two coordinate transformations 
corresponding to the two extensions,
%
%
\begin{displaymath}
\phi'=\phi\pm{1\over2\sqrt{M}}
\log{\sqrt{M}+\sqrt{M-\rho/l^2} \over 
\sqrt{M}-\sqrt{M-\rho/l^2}}\ \ .
\end{displaymath}
They give the two expressions for the metric,
%
%
\begin{equation}
ds^2=\pm{d\phi'd\rho \over \sqrt{M-\rho/l^2}} + \rho d\phi'^2 
+\left(M-\rho/l^2\right) dt^2\ \ ,
\end{equation}
respectively.
Since these metrics are not degenerate on $\rho=0$, we can extend the 
spacetime 
from the region with $\rho>0$ to the region with $\rho<0$. 
By one of the two extensions a certain set of incomplete geodesics become 
complete, while the other 
extension makes the other set of incomplete geodesics complete.

The identification 
$(\phi\leftrightarrow\phi + 2\pi)$ should exist also in the extended 
region. The action of the transformation $T$ (see (\ref{eqn:mat})) in the 
extended region 
generates this identification. Fig.2 shows this identification in the 
original 
 region and also in the extended region for a 
$x, z=constant$ section of the pseudo-sphere $x^2+y^2-z^2-w^2=l^2$. Thus 
we find that the topology of the space changes from a single $R\times 
S_1$ (the original region) to two $R\times S_1$'s (the extended region).
It is noted that the spacetime is non-Hausdorff on the boundary between 
the original region and the extended region $y^2-w^2=0$, if we do both 
of the extensions.
%
%
\begin{figure}[h]
\centerline{\epsfxsize=10.0cm \epsfbox{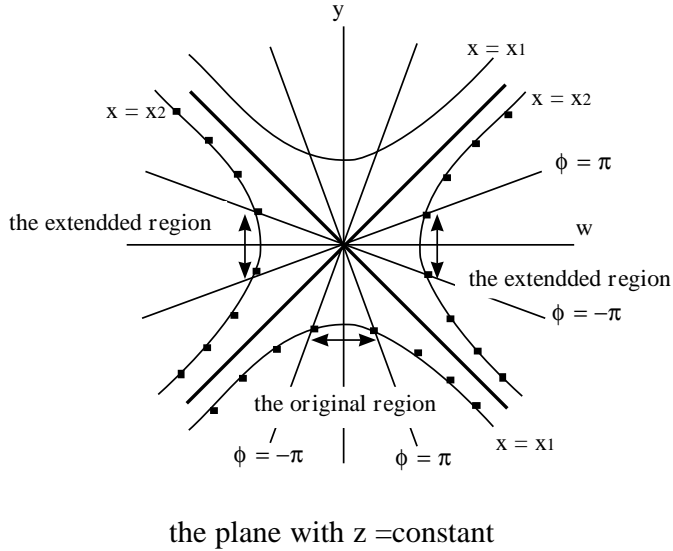}}
\caption[A sectiion]{A section of the pseudo-sphere with $x, z=constant$ is a 
hyperbola 
parametrized by $\phi =1/ \sqrt{M} \tanh^{-1}(w/y)$. A point
$\phi =\phi_0$ is identified with a point $\phi=\phi_0 + 2\pi$.}
\end{figure}

\subsection{Compactified 3D-BH}
\hspace*{\parindent}
We extend also the compactified 3D-BH by the above mentioned method. 
The original region is the region with $x^2<l^2$; it is easily confirmed 
that this region coincides with the region covered by the RW-chart. The 
coordinate extension is 
carried out in a similar way as the ordinary 3D-BH. This time there are 
eight 
extensions corresponding to the eight directions of the identification. By 
considering the covering space of the RW-anti-de Sitter spacetime, we find 
that this extended region is the region  $x^2-l^2=z^2+w^2-y^2>0$ of the 
3-anti-de Sitter not 
covered by the RW-chart (see Fig.3).
%
%
\begin{figure}[h]
	\centerline{\epsfxsize=12.0cm \epsfbox{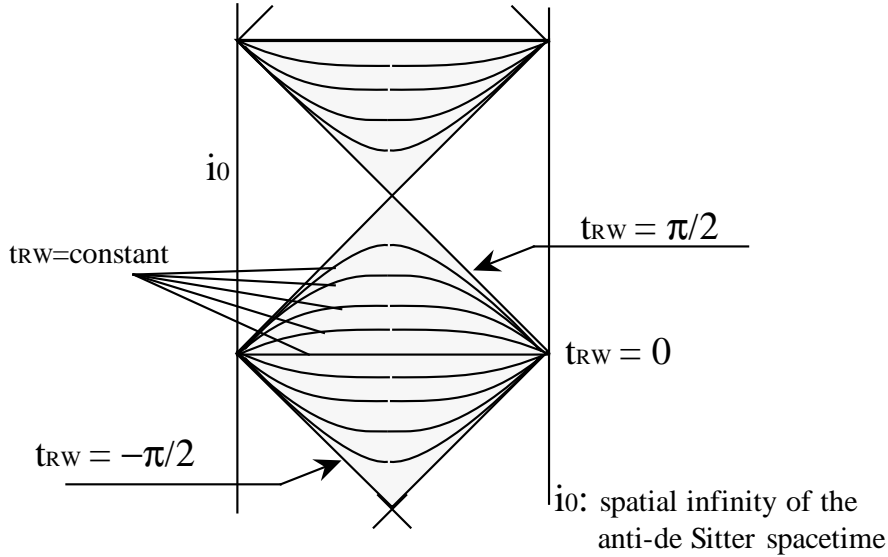}}
	\caption{The conformal diagram of the anti-de Sitter spacetime. The 
	RW-chart covers only a part of the whole anti-de Sitter spacetime, as 
shown
by the shadowed region.}
\end{figure}
The original region is foliated by a one parameter family of 2-hyperbolic 
surfaces:
%
%
\begin{displaymath}
H_2(x_0^2)=\{y^2-z^2-w^2=\lambda^2\equiv l^2-x_0^2>0\} ,
\end{displaymath}
with $x_0^2(<l^2)$ being a parameter.
The identifications shown in Fig.1 produce a double 
torus at each surface. As stated in the previous section, these 
identifications are 
defined by a discrete subgroup of $SO(2,2)=$Isom(3-anti-de 
Sitter) generated by $T_i$ in (\ref{eqn:dt}).

We also make such identifications 
in the extended region in the following way. 
The extended region is foliated by the (1+1)-de 
Sitter spacetimes:
%
%
\begin{displaymath}
dS_2(x_0^2)=\{z^2+w^2-y^2=\lambda '^2\equiv x_0^2-l^2>0\} ;
\end{displaymath}
parametrized by $x_0^2(>l^2)$.
Since the discrete subgroup preserves each $dS_2(x_0^2)$, the 
identifications are induced for each hypersurface. We can find the 
topology of this 
region by determining the fundamental region in the hypersurface 
$dS_2(x_0^2)$.
The action of $T_1$ on the surfaces $dS_2(x_0^2)$ is shown in Fig.4.
%
%
\begin{figure}[h]
	\centerline{\epsfxsize=10.0cm \epsfbox{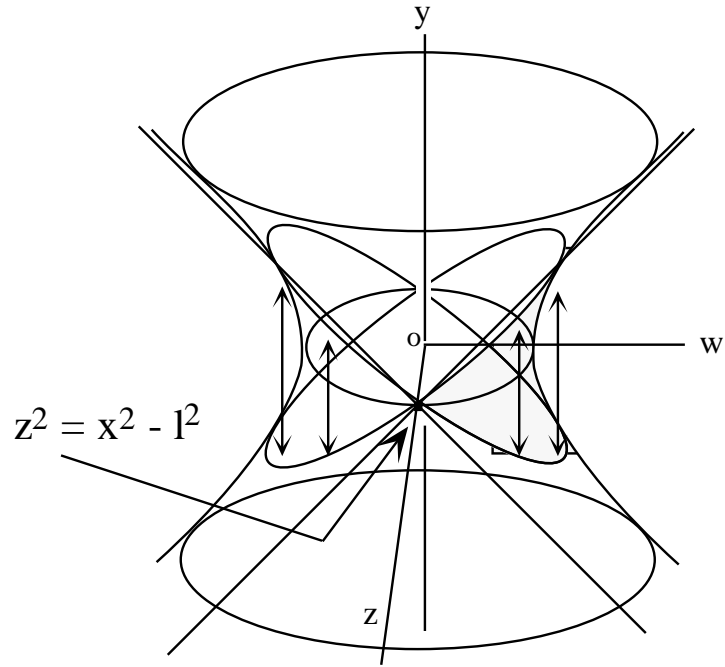}}
	\caption{The action of $T_1$ on a $dS_2(x_0^2)$ is shown by the arrows.}
\end{figure}
The other actions of transformations are given by a rotation 
of $T_i$ by $\pi (1-i)/4$ with $i=1\sim 4$. The fundamental regions are 
shown in Fig.5 both for $H_2(x_0^2)$ and for 
$dS_2(x_0^2)$.
%
%
\begin{figure}[h]
	\centerline{\epsfxsize=11.0cm \epsfbox{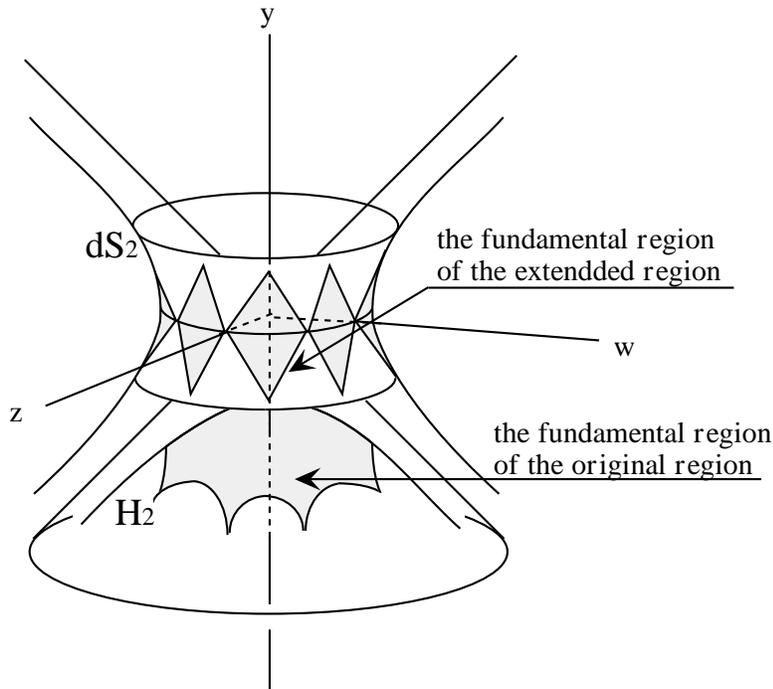}}
	\caption{While the fundamental 
region is an octagon, in the $H_2(x_0^2)$ region of spacetime it changes 
to eight quadrilaterals in $dS_2(x_0^2)$.}
\end{figure}
The fundamental region of the $dS_2(x_0^2)$ consists of eight 
quadrilaterals. Fig.6 shows 
that such a quadrilateral is topologically equivalent to a sphere with 
three conical singularities.

To summarize, the analytical continuation of 
the compactified 3D-BH produces the topology changing geometry from a 
double torus to eight spheres with three punctures. This topology change 
is quite different from the quantum topology change investigated in 
Ref.\cite{FHHMS}. Previously the quantum topology change was 
described by a Euclidean tunneling manifold whereas the present 
topology changing solution has the Lorentzian signature. The 
topology change presented in this paper 
happens in classical gravity.
We also note that the 
spacetime is non-Hausdorff, and it is discussed in detail in 
\cite{TH}. As stated in \cite{HE}, this kind of non-Hausdorff topology is 
not so uncomfortable as the case shown in Fig.8, since we have no 
continuous curve which bifurcates.  
%
%
\begin{figure}[h]
	\centerline{\epsfxsize=12.0cm \epsfbox{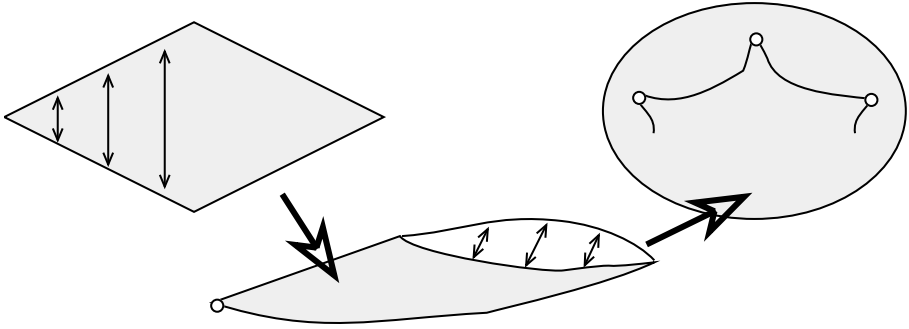}}
	\caption{The quadrilateral becomes a sphere with three punctures by the 
	identifications.}
\end{figure}

\section{Closed Geodesic Curve}
\hspace*{\parindent}
Generally an identification of points creates new closed curves; these 
closed curves link the two identified points on the boundary of the 
fundamental region. Since, in the case of 
the Misner 
solution, such closed curves become closed timelike curves in its 
extended region, one can naturally expect that the ordinary 3D-BH and 
the compactified 3D-BH contain at least a closed timelike curve in their 
extended 
regions. 

To see this more precisely we consider the homotopy class of smooth curves 
emanating 
from 
one point on the boundary of the fundamental region and ending at the 
point 
identified with it, on each hypersurface (see Fig.7). If this homotopy 
class contains a curve which is 
everywhere timelike (spacelike), a closed timelike (spacelike) curve 
exists and 
passes through the point on the boundary of the fundamental region. As the 
metric
is continuous, when each homotopy class contains its own everywhere 
timelike 
(spacelike) curve, the homotopy classes contain no everywhere spacelike 
(timelike) curve. 
Because all non-trivial curves are homotopic to the 
combination of these closed curves crossing the boundary of the fundamental 
region once, we can see the causal property of 
the all non-trivial closed curves, finding out the 
everywhere 
spacelike or everywhere timelike curves.
%
%
\begin{figure}[h]
	\centerline{\epsfysize=12.0cm \epsfbox{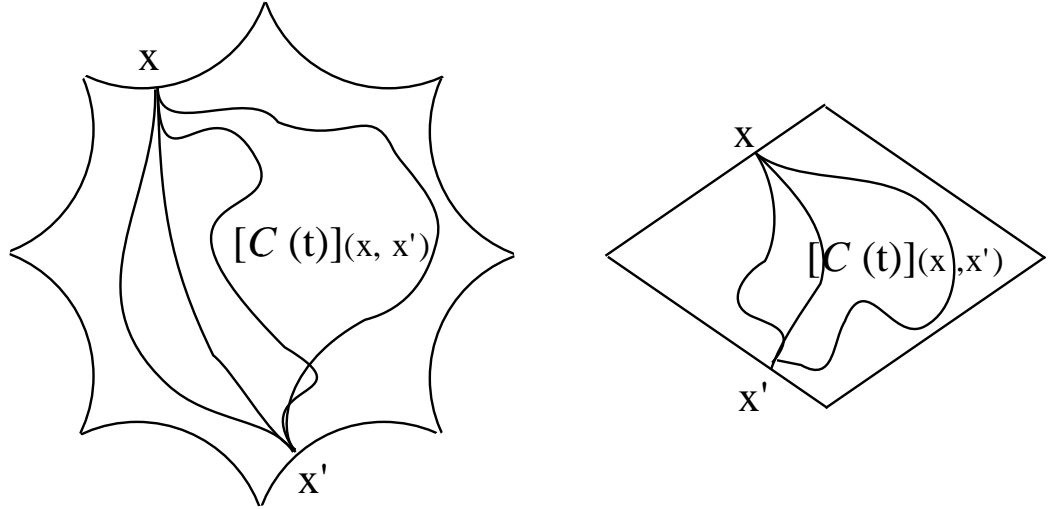}}
	\caption{A homotopy class of smooth curves emanating 
from 
one point on the boundary of the fundamental region and ending at the 
point 
identified with it.}
\end{figure}
 
Here we consider a curve which is an orbit of an element of the discrete 
subgroup bearing in mind that
this curve is  smooth even after the identification.
In particular, if we vary the boost angle of a Lorentz transformation $T_i$, this one 
parameter family of transformation 
generates a smooth curve. For example, the Lorentz boost (\ref{eqn:mat}) generates 
such a curve,
%
%
\begin{equation}
\label{eqn:ct}
{\cal C}(t)=
\left(\begin{array}{cccc}
1 & 0 & 0 & 0 \\
0 & \cosh t & 0 & \sinh t \\
0 & 0 & 1 & 0 \\
0 & \sinh t & 0 & \cosh t
\end{array}\right)
\left(\begin{array}{c}
x \\
y \\
z \\
w 
\end{array}\right) ,
\end{equation}
with $t$ being a boost angle.
The expression (\ref{eqn:ct}) can be rewritten as 
%
%
\begin{equation}
\label{eqn:exp2}
{\cal C}(t)=exp\left[t
\left(\begin{array}{cccc}
0 & 0 & 0 & 0 \\
0 & 0 & 0 & 1 \\
0 & 0 & 0 & 0 \\
0 & 1 & 0 & 0
\end{array}\right)\right] 
\left(\begin{array}{c}
x \\
y \\
z \\
w 
\end{array}\right) .
\end{equation}

On the other hand, the integral curve of the vector field
$X^a$ is generally given by the exponential map,
%
%
\begin{equation}
\label{eqn:exp1}
\gamma(s)=exp_q(s X^a) ,
\end{equation}
so that we obtain the tangent vector to the curve ${\cal C}(t)$ given in 
Eq.(\ref{eqn:exp2}) as
%
%
\begin{equation}
X^a
=\left(\begin{array}{c}
0 \\
w \\
0 \\
y \\
\end{array}\right) .
\end{equation}
The square $X^aX_a$ is $-w^2+y^2$.

In the ordinary 3D-BH closed non-trivial curves are 
homotopic to the curves ${\cal C}(t)$'s with fixed starting points. 
In the original region of the ordinary 3D-BH ($y^2-w^2>0$), as $X^a$ is 
always spacelike 
there is a closed non-trivial spacelike curve but no closed timelike 
curve. On the contrary the extended region ($y^2-w^2<0$) contains a closed 
non-trivial timelike curve but no closed spacelike curve as we shall see 
below. On the boundary 
surface ($y^2-w^2=0$) such closed curves become null. 

For the identifications of the compactified 3D-BH, the calculations are 
similar. We determine the tangent vectors $X^a[T_i]$'s as follows,
%
%
\begin{eqnarray}
X^a[T_1] &=&
\left(\begin{array}{c}
0 \\ w \\ 0 \\ y
\end{array}\right) \ \ \ ,\ \ \ X^aX_a=-w^2+y^2 \\
X^a[T_2] &=&{1\over \sqrt{2}}
\left(\begin{array}{c}
0 \\ z+w \\ y \\ y
\end{array}\right) \ \ \ ,\ \ \ X^aX_a=-{(w+z)^2 \over 2}+y^2, \\
X^a[T_3] &=&
\left(\begin{array}{c}
0 \\ z \\ y \\ 0
\end{array}\right) \ \ \ ,\ \ \ X^aX_a=-z^2+y^2, \\
X^a[T_4] &=&{1\over \sqrt{2}}
\left(\begin{array}{c}
0 \\ -z+w \\ -y \\ y
\end{array}\right) \ \ \ ,\ \ \ X^aX_a=-{(w-z)^2 \over 2}+y^2.
\end{eqnarray}
 
In the original region of the compactified 3D-BH $X^a[T_i]$ is 
everywhere spacelike because $y^2-w^2-z^2=l^2-x^2$ is positive there. Then 
each homotopy class of closed curves contains a closed spacelike curve 
but no closed timelike curve. The case of the extended region of the 
compactified 
3D-BH  ($y^2-w^2-z^2<0$) is not so straightforward so that we need closer 
examination of the curves. For instance, we 
consider the curve generated by $T_1$. $X^a[T_1]$ is spacelike only in 
the region $z^2> \lambda'^2\equiv l^2-x_0^2$. The fundamental region of 
the 
extended region, however, is contained in the region where 
$-z^2+\lambda'^2$ is negative as depicted in Fig.4. Therefore $X^a[T_1]$ 
is everywhere 
timelike in the 
fundamental region of the 
extended region of the compactified 3D-BH. As this aspect is the same for 
all the other $T_i$'s, we can conclude that the extended 
region contains non-trivial closed timelike curves corresponding to each 
Lorentz boost $T_i$ and therefore there is no 
non-trivial closed spacelike curve. 
\section{Summary and Discussions}
\hspace*{\parindent}
In this paper, we have investigated the analytical continuations of the 
ordinary 3D-BH and of the compactified 3D-BH. The compactified 
3D-BH is isometric to a regular double torus universe with a negative 
cosmological constant if the mass 
of the black hole is equal to $m_c=({1 / \pi 
}\tanh^{-1}\sqrt{(2\sqrt{2}-2)/3})^2$. If the mass of the black hole were 
different from $m_c$, however, the double torus would 
contain a conical singularity. Even if so, all the results of this paper 
will not be affected. 

One might argue that the compactified 3D-BH would not be a proper black 
hole. Any compactification would remove the spatial infinity 
and the null infinity 
from the spacetime. Then, as the argument may go, we could not define any 
event horizon. In this 
paper, however, we treat the compactified 3D-BH solution as a black hole 
because the double 
torus universe can be regarded as an open 2-hyperbolic space which 
possesses 
a certain periodicity. The horizon can be defined in the universal covering 
spacetime. To summarize we can define a compactified space as a black hole if 
its universal covering space is a black hole spacetime.

 We have shown that the analytical continuations of the 3D-BH's 
produce a topology changing spacetime geometry. Especially the topology change 
of the ordinary 3D-BH is easily attributed to 
the singularity where the spacetime is non-Hausdorff. A well-known simple 
example 
of non-Hausdorff topological space is shown in Fig.8\cite{HE},
and is the simplest example of a topology change, from one 
point to two points, if one imagines that the $x$ and $y$ axes are the time 
axes. The non-Hausdorff topology causes the branching of 
the space which is one type of the topology change. In the ordinary 
3D-BH, the topology of a hypersurface changes from 
a single $R\times S_1$ to two $R\times S_1$'s. This topology change 
belongs to the branch-type 
topology change corresponding to the change of the number of connected 
components. 
However, the topology change 
of the compactified 3D-BH (a double torus universe) is not so trivial,
since the Euler number changes from $-2$ (a double torus) to $8\times 
(-1)$ (eight spheres with three punctures). This can be seen by 
considering the geodesics 
running from the original region to the extended region\cite{TB}. Anyway 
these topology change is closely related to non-Hausdorff topologies. 
We may speculate that all Lorentzian topology changes are related to the 
non-Hausdorff topology. 
%
%
\begin{figure}
	\centerline{\epsfxsize=11.0cm \epsfbox{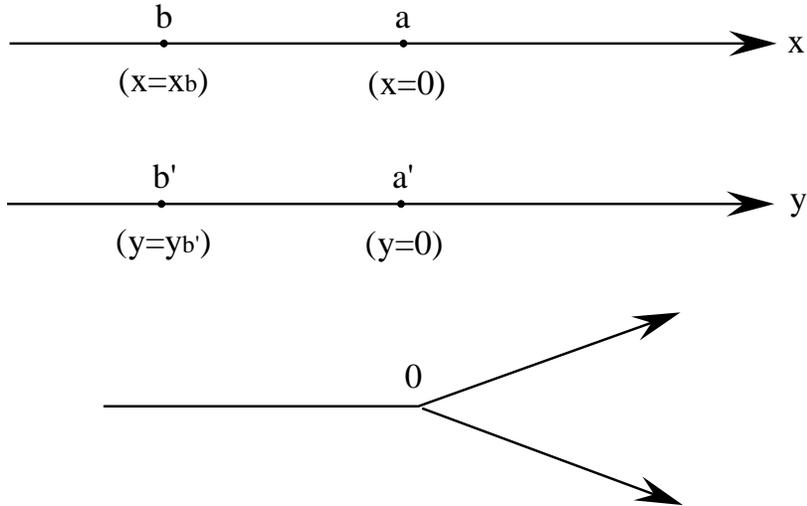}}
	\caption{An example of a non-Hausdorff manifold. The two lines above are 
	identical for $x=y<0$. However, the two points $a$ ($x=0$) and $a'$ (y=0) 
	are not identified.}
\end{figure}

As for the problem of the topology change in a Lorentzian spacetime, 
some authors have discussed its possibility or the condition for it to 
occur\cite{GST}\cite{GH}. 
These works, however, considered spacelike hypersurfaces as an initial 
and a final states. As 
shown in this paper there is a new type of
topology change if we consider timelike hypersurfaces.

The investigation of this paper about a double torus is also applicable to 
a 2-surface 
with higher genus. When we consider the case of genus $n>2$, the mass of 
the black hole should be $(1/\pi\tanh^{-1}\sqrt{2/(1+\sec 
(\pi/2n)})^2$ to form a regular spatial hypersurface with a genus 
$n$. In this case the topology changes from a single $n$-torus to 4$n$ 
spheres with three conical singularities.

 The existence of the closed 
timelike curve means the break down of causality. Hawking conjectured that 
such a closed timelike curve cannot be formed since the energy momentum 
tensor for quantum matter fields 
diverges near the cauchy horizon which should appear if the closed 
timelike curve is created\cite{CP}. This problem is still an 
open 
question\cite{CP}\cite{TR}. The spacetime shown in this paper may be an 
appropriate 
background spacetime to discuss this problem.

For the open chart of 3-de Sitter spacetime or the 3-Milne
universe, the same compactification and extension can be done. In these cases we will also find a topology change 
and a closed timelike curve. It is our future plan to study the cosmological
implication in the 
Friedmann spacetimes with compactified hyperbolic spatial hypersurface. 
Furthemore we can study the Teichm\"{u}ller deformation and its quantum 
theory of a further simplified model, for example a torus with a conical 
singularity in the 3D-BH\cite{TB}.  
\begin{flushleft}
\Large{\bf Acknowledgments}
\end{flushleft}

I would like to thank Professor A. Hosoya for helpful discussions. I am 
grateful to Professor H. Ishihara for useful discussions.
The author thanks the Japan Society for the 
Promotion of Science for financial support. This work was supported in 
part 
by the Japanese Grant-in-Aid for Scientific Research Fund of the Ministry 
of 
Education, Science and Culture.



\begin{thebibliography}{99}

\bibitem{FHHMS}
Y.~Fujiwara, S.~Higuchi, A.~Hosoya, T.~Mishima and M.~Siino, {\it Phys. 
Rev. \bf D44 \rm (1991) 1756}, {\it Phys. 
Rev. \bf D44 \rm (1991) 1763}, {\it Class. Quantum Grav. \bf 7 \rm (1992) 
163}.

\bibitem{GST}
R.~P.~.Geroch, {\sl J. Math. Phys. \bf 8 \rm(1967) 782}, P.~Yodzis, {\sl 
Gen. Rel. Grav. \bf 4 \rm(1973) 299}, D.~Gannon, {\sl J. Math. Phys. \bf 
16 \rm (1975) 2364}, F.~J.~Tipler, 
{\sl Ann. Phys. \bf 108 \rm (1977) 1}, R.~D.~Sorkin, {\sl Phys. Rev. \bf 
D33 \rm (1986) 978}. 

\bibitem{HNM}
A.~Hosoya and K.~Nakao, {\it Class. Quantum Grav. \bf 7 \rm (1990) 
163} ,{\it Prog. Theor. Phys. \bf 84 \rm(1990) 739},
V.~Moncrief, {\it J. Math. Phys. \bf 30 \rm (1989) 2907}. 



\bibitem{DJT}
S.~Deser, R.~Jackiw and G.~'tHooft, {\it Ann. of Phys. \bf 152 \rm (1984) 
220}.



\bibitem{BTZ}
M.~Ba\~{n}ados,C.~Teitelboim and J.~Zanelli, {\it Phys. Rev. Lett. \bf 69 
\rm(1992) 1849}.

\bibitem{MI}
C.~W.~Misner {\sl~Relativity Theory and Astrophysics I: Relativity and 
Cosmology, \rm ed. J.~Ehlers, Lectures in Applied Mathematics, Volume 8} 
(American Mathematical Society, 1967) 160-9.

\bibitem{HE}
S.~W.~Hawking and G.~F.~R.~Ellis, {\sl~The large 
scale structure of space-time} (Cambridge University
 Press, Cambridge, 1973).
 

\bibitem{GH}
G.~W.~Gibbons and S.~W.~Hawking, {\sl Commun. Math. Phys. \bf 148 
\rm(1992) 345}. 

\bibitem{TH}
M.~Siino, Ph.D.~Thesis, T.I.Tech. (1994).

\bibitem{TB}
M.~Siino,  in preparation.

\bibitem{CP}
S.~W.~Hawking {\sl Phys. Rev. \bf D46 \rm(1992) 603}.

\bibitem{TR}
M.~S.~Morris, K.~S.~Thorne and U.~Yurtsever, {\sl Phys. Rev. Lett. \bf 61 
\rm(1988) 1446}.
\end{thebibliography}
\end{document}